\documentclass[10pt]{article}

\usepackage{amsmath}
\usepackage{amsfonts}
\usepackage{graphicx}

\begin{document}

\title{\Large\bf Perfect Fluids: Field-theoretical Description
and Gauge Symmetry Issue}

\author{\bf Nikolai V. Mitskievich\thanks{Departamento de
F{\'\i}sica, C.U.C.E.I., Universidad de Guadalajara, Guadalajara,
Jalisco, M\'exico. \hspace*{1.5cm}E-mail: nmitskie@gmail.com}}

\maketitle

\begin{abstract}
We show that combinations of (in general, non-linear) 2- and
3-form fields analogous to the Maxwell (1-form) field, completely
describe perfect fluids, including the rotating ones. In the
non-rotating case, the 2-form field in sufficient, and a free
3-form field proves to be equivalent to appearance of the
cosmological term in Einstein's equations (the square-root
non-linearity corresponding to $\Lambda=0$). The gauge degrees of
freedom break down when a rotation is included, but even when they
exist, there obviously fails to be realized an equivalence of the
2-form field and the massless scalar one recently claimed by
Weinberg.

\hrulefill
\end{abstract}

\vspace*{1cm}

We consider here $r$-form fields ($r=$ 0, 1, 2 and 3, the
corresponding $r$-forms for potentials being
$\varphi$, $A$, $B$, and $C$). Let the Lagrangian density
${\frak L}=\sqrt{-g}L$ be a function of Maxwell-type
invariants ($H=\ast(\Phi\wedge\ast\Phi)$, {\it etc.} for
$I$, $J$ and $H$) of the corresponding
field tensors, the $(r+1)$-forms $\Phi=d\varphi$, $F=dA$,
$G=dB$, and $H=dC$. Thus the Lagrangian depends on
the metric coefficients only algebraically. We use
the spacetime signature $+$, $-$, $-$, $-$,
and 4-dimensional Greek indices, $\ast$ being the Hodge star.
Below the dependence on Maxwell's field $F$ will be omitted.

The 2nd Noether theorem (see \cite{Mitskievich58}) gives in this
case the standard definition of stress-energy tensor
\begin{equation}
{\frak T}^{\mu\nu}=\sqrt{-g}T^{\mu\nu}=
-2\frac{\partial{\frak L}}{\partial g_{\mu\nu}}
\end{equation}
(only the variational derivative with respect to $g_{\mu\nu}$
is reduced to the partial one), so that
\begin{equation}
T=-Lg+2H\frac{\partial L}{\partial H}\stackrel{\scriptsize (0)}{u}
\times\stackrel{\scriptsize (0)}{u}+2J\frac{\partial L}{\partial J}
\left(g-\stackrel{\scriptsize (2)}{u}\times
\stackrel{\scriptsize (2)}{u}\right)+
2K\frac{\partial L}{\partial K}g        \label{T}
\end{equation}
where $\stackrel{\scriptsize (0)}{u}=\Phi/\sqrt{-H}$
and $\stackrel{\scriptsize (2)}{u}=\tilde{G}/\sqrt{J}$,
while $\tilde{G}_\alpha=(1/3!)G^{\lambda\mu\nu}
E_{\lambda\mu\nu\alpha}$; $E_{\kappa\lambda\mu\nu}:
=\sqrt{-g}\epsilon_{\kappa\lambda\mu\nu}$,
$\epsilon_{\kappa\lambda\mu\nu}$ being
the Levi-Civit{\`a} symbol.

The phenomenological stress-energy tensor of a perfect fluid has
in our notations the form $T^{\rm pf}=(\mu+p)u\otimes u-pg$. Here
$p$ is invariant pressure of the fluid, $\mu$ its invariant mass
(energy) density, and $u$ its local four-velocity. Thus a
stress-energy tensor acceptable for description of perfect fluids,
should have one distinct eigenvalue $\mu$ corresponding to the
eigenvector $u$ and another (now, triple) eigenvalue $(-p)$
corresponding to any vector of the whole local subspace orthogonal
to $u$. One may find information about the main stages of
development of the theory of perfect fluids in \cite{Taub54},
\cite{Schutz70}, \cite{Brown93}, \cite{Carter94}; this paper was
first published in \cite{Mitskievich99} in a somewhat shorter
form; also see a much more detailed paper [9] (though without
gauge-related considerations).

Let us first consider the pure free field case: only one
of the $r$-form fields should be then present at once, or
$L=\stackrel{\scriptsize (0)}{L}+\stackrel{\scriptsize (2)}{L}+
\stackrel{\scriptsize (3)}{L}$, the consecutive terms depending
on the corresponding $r$-form field variables.
The tensor (\ref{T}) is compatible with the above conditions
for the cases $r=$ 0, 2 or 3 only (Maxwell's field
and its non-linear analogues do not meet the requirements,
hence they were already omitted).
A comparison with (\ref{T}) yields
$\stackrel{\scriptsize (0)}{\mu}=2H\,\partial\!
\stackrel{\scriptsize (0)}{L}/\partial H-\!
\stackrel{\scriptsize (0)}{L}$,  $\stackrel{\scriptsize (0)}{p}=
\stackrel{\scriptsize (0)}{L}$; $\stackrel{\scriptsize (2)}{\mu}
=-\!\stackrel{\scriptsize (2)}{L}$,
$\stackrel{\scriptsize (2)}{p}=\stackrel{\scriptsize (2)}{L}-
2J\,\partial\!\stackrel{\scriptsize (2)}{L}/\partial J$;
$\stackrel{\scriptsize (3)}{\mu}=-\stackrel{\scriptsize (3)}{p}
=-\!\stackrel{\scriptsize (3)}{L}+2K\,\partial\!
\stackrel{\scriptsize (3)}{L}/\partial K$.
It is also clear that the vector $\stackrel{\scriptsize (0)}{u}$
is timelike (thus suitable for description of four-velocity),
only if the scalar field $\varphi$ is essentially non-stationary,
but for $\stackrel{\scriptsize (2)}{u}$ there is the exactly
opposite situation: in the 2-form field potential the $t$-dependence
has not to dominate, or it even may be absent (for a static
or stationary 2-form field). Moreover, the $p=0$ case (incoherent
dust) cannot be described at all via the scalar field, in contrast
to the 2-form field. Thus one has to exclude the 0-form
(massless scalar) field if the problem under consideration is
to describe a perfect fluid which can be brought to its rest frame
(at least locally). Therefore the superscript {\scriptsize (2)}
in $\stackrel{\scriptsize (2)}{u}$ will henceforth be omitted.

In the case of a pure 3-form field, the stress-energy tensor (\ref{T})
is explicitly proportional to $\delta^\beta_\alpha$: it is equivalent
to addition of a cosmological term to Einstein's equations. From
(\ref{T}) it is also obvious that $T^\beta_\alpha$ identically
vanishes when $L\sim\sqrt{K}$. This latter case can be called that of
a phantom 3-form field which may be described by an {\em arbitrary}
function of coordinates. In the former case, the 3-form field is
simply constant (we speak on only one function since everything is
determined here by a pseudoscalar, the dual conjugate to the 4-form
$H$). The both cases follow also from the field equations being a
result of variational principle applied to $L$; though the 3-form
field is non-dynamical in this sense, it affects the global
geometry of the universe via determination of the cosmological term,
and it may provide virtual particles in quantum theoretical
Feynman-type graphs, when coupled to other fields. Thus one could
relate this field to the fundamental cosmological field proposed
by Sakurai \cite{Sakurai60} (another reason is its decisive role in
description of rotating fluids, especially when one thinks about
the global aspects of rotation and the Mach principle).

Weinberg \cite{Weinberg96} has given a generalization of the gauge
field theory (essentially of the electromagnetic field) to the
case of $p$-form fields (Section 8.8 of his very instructive and
well written book; we have to speak here about the $r$-form fields
simply because $p$ means pressure of the fluid in our context).
Weinberg's main conclusion in this respect was that `in four
spacetime dimensions, $p$-forms offer no new possibilities': $p=3$
is simply an empty case, and $p=2$ `is equivalent to a zero-form
gauge field, which as we have seen is equivalent to a massless
derivatively coupled scalar field'. Our communication however
represents a counterexample to these conjectures, as it is seen
from the last two paragraphs above. The error committed by
Weinberg was that the physical interpretation of $p$-form fields
(as well as all other geometric objects playing r\^oles of
physical fields) does not reduce to their geometric properties,
but it crucially depends on their dynamical equations, {\it i.e.}
the structure of corresponding Lagrangians.

In the pure 2-form field case, it is easy to translate into
the field theoretical language all
general relativistic solutions for non-rotating fluid
(for all cases of linear dependence of $p$ on $\mu$, as well
as for polytrope equations of state; the only known exception
is the interior Schwarzschild solution which can be translated
in the context of interacting 2- and 3-form fields).

In the case of a pure 2-form field, the field equations reduce
to
\begin{equation}
d\left(J^{1/2}\frac{dL}{dJ}u\right)=0,   \label{2a-eq}
\end{equation}
$u$ being the normalized 1-form of $\tilde{G}$ (four-velocity
of the fluid). One finds immediately that the fluid does not
rotate. The only remedy is
in this case an introduction of a source term
in (\ref{2a-eq}).

The simplest way to do this is to introduce in the Lagrangian
density dependence on a new invariant $J_1=
-B_{[\kappa\lambda}B_{\mu\nu]}B^{[\kappa\lambda}B^{\mu\nu]}$
which does not spoil the structure of stress-energy tensor
(alongside with $J_1$, we shall use the old invariants $J$
and $K$). Since
\begin{equation}
B_{[\kappa\lambda}B_{\mu\nu]}=-\frac{2}{4!}B_{\alpha\beta}
B\!\stackrel{\alpha\beta}{\ast}E_{\kappa\lambda\mu\nu}
\end{equation}
where $B\!\stackrel{\alpha\beta}{\ast}:=\frac{1}{2}B_{\mu\nu}
E^{\alpha\beta\mu\nu}$ (dual conjugation),
$J_1^{1/2}=6^{-1/2}B_{\alpha\beta}
B\!\!\stackrel{\alpha\beta}{\ast}$.
In fact, $J_1=0$, if $B$ is a simple bivector (this
corresponds to all types of rotating
fluids discussed in existing literature). This does not
however annul the expression which this invariant
contributes to the 2-form field equations: it is proportional
to $\partial J_1^{1/2}/\partial B_{\mu\nu}\neq 0$.
Thus let the Lagrangian density be
\begin{equation}
{\frak L}=\sqrt{-g}\left(L(J)+M(K)J_1^{1/2}\right).
\end{equation}

The 2-form field equations now take the form
\begin{equation}
\displaystyle d\left(\frac{dL}{dJ}\tilde{G}\right)=
\sqrt{2/3}M(K)B.   \label{2-eq}
\end{equation}
In their turn, the 3-form field equations
yield the first integral
\begin{equation}
J_1^{1/2}K^{1/2}\frac{dM}{dK}=\mbox{const}\equiv 0
\end{equation}
(since $J_1=0$) in agreement with the fact
that $K$ (hence, $M$) is an {\em arbitrary} function,
if only the 3-form field
equations are taken into account. Though the
equations (\ref{2-eq}) apparently show that the $\tilde{G}$ congruence
should in general be rotating, the 2-form field $B$ is an exact
form for solutions with constant $M(K)$, thus its substitution
into the left-hand side of (\ref{2-eq}) via $\tilde{G}$, leads
trivially to vanishing of $G$ (and hence $B$). We see that in
a non-trivial situation the cosmological field $K$
has to be essentially non-constant.

But the complete set of equations contains Einstein's
equations as well. One has to take into account their sources and
the structure of their solutions
in order to better understand this remarkable
situation probably never encountered in theoretical physics
before.

The gauge freedom suggested by $G=dB$, is obviously destroyed
by the field equations (\ref{2-eq}) when a rotation is switched
on. Since the rotation is so widespread in nature, existence
of the gauge freedom in $B$ should merely be an exclusion
than a rule.

\section*{\large\bf ACKNOWLEDGEMENTS}

This work was supported by the CONACyT Grant 1626P-E,
by a research stipend of the Albert-Einstein-Institut
(Potsdam), and by a travel grant of the Universidad de
Guadalajara. My thanks are due to B. Carter, J. Ehlers,
F.W. Hehl and H. Vargas Rodr{\'\i}guez for valuable
information and friendly discussions.

\end{document}